\documentclass[amssymb,aps,floats,prb,twocolumn]{revtex4}
\usepackage{color,graphicx,pstricks,subfigure}
\usepackage{bm}
\usepackage[dvips]{rotating}
\usepackage{epstopdf}
\usepackage[abs]{overpic}

\makeatletter\renewcommand\@biblabel[1]{{#1}}\makeatother

\begin{document}

\title{\,
Nodes to the grindstone:\\
    viewpoint on ``Band- and momentum-dependent electron dynamics in
    superconducting Ba(Fe$_{1-x}$Co$_x$)$_2$As$_2$ as seen via electronic Raman
    scattering"
}
\author{{}
P.~J.~Hirschfeld \vspace{0cm}}

\affiliation{Department of Physics, University of Florida,
Gainesville, FL 32611, USA \vspace{0cm} } \vspace{0cm}
\date{\today}\vspace{0cm}

\begin{abstract}

\end{abstract}

\maketitle

Materials become superconductors when charge carriers pair up; this pairing
is stabilized by an energy gap, the energy cost for creating single-particle excitations.
In any new class of superconducting materials, the momentum space
structure of this energy gap (or superconducting order parameter)
 is generally considered to be one
of the most important clues to the nature of the pairing
mechanism. Determining the order parameter is rarely
straightforward, and historically in, e.g. the cuprates, many
different types of experiments had to be analyzed and compared
before a consensus was achieved. It is therefore not extremely
surprising that, nearly two years after the discovery of the high
temperature Fe-pnictide superconductors, the symmetry and form of
the order parameter in these systems are still controversial.
Still, the degree of apparent disagreement among different
experiments on similar samples\cite{reviews} has raised the
question: can the superconducting state of these materials be
extraordinarily sensitive to either disorder or other aspects of
electronic structure which ``tune" the pairing interaction?

Based on density functional theory, quantum oscillations and
angle-resolved photoemission experiments (ARPES), the Fermi
surface of the Fe-pnictides is thought to consist of
 of a few small hole and electron pockets.  Several experiments
have been interpreted in terms of order parameters which are
isotropic (independent of momentum on a given pocket), but
possibly with overall sign change between electron and hole
pockets as predicted by theory\cite{splusminus}.
 On the other hand, many experiments have indicated the
existence of low-lying excitations below the apparent gap energy.
A natural way of interpreting these observations is to assume that
the order parameter has nodes on some part of the Fermi surface,
such that quasiparticles can be created at arbitrarily low
energies.  An alternative explanation for this second set of
experiments has been proposed, however; in an isotropic
``sign-changing $s$-wave" ($s_\pm$) superconductor (Fig. 1a),
disorder can create subgap states\cite{disorder} under certain
conditions, depending on the ratio of inter- to intraband impurity
scattering. From the theoretical standpoint, the most likely
states indeed appear to be preferentially of "$s$-wave" symmetry,
with quasi-isotropic gaps on the hole pockets but potentially
highly anisotropic (nodal, e.g. Fig. 1b or near-nodal, Fig. 1d)
states on the electron pockets\cite{sf_theory}. At present it is
not completely clear from these theories what drives the
anisotropy on the electron pockets, although some useful
observations have been made\cite{anisotropy}.
 It is clearly extremely important
to establish empirically whether low-energy excitations are
intrinsic (nodal) or extrinsic (disorder-induced), and under what
circumstances fully developed gaps should be expected.

In this regard, electronic Raman scattering in the superconducting
state is an ideal probe.  In addition to sensitivity to low-lying
excitations, Raman scattering can be performed for various
polarizations of the incoming and outgoing photons, so as to
preferentially sample the excitations created in different parts
of the Brillouin zone.  In a recent article, Muschler et
al.\cite{Muschler} have presented Raman scattering measurements on
single crystals of Ba(Fe$_{1-x}$Co$_x$)$_2$As$_2$ for two
different (near optimal and slightly overdoped) concentrations of
Co.  Similar measurements and theoretical interpretations were
instrumental early on in identifying $d$-wave pairing in the
cuprates, but the scattering from charge excitations  is up to an
order of magnitude smaller in the Fe-pnictide systems, so the mere
observation of a change in the signal with temperature below $T_c$
by Hackl and co-workers is a significant achievement; it indicates
that Raman experiments can again play an important role in
determining the symmetry of the new materials.

 Muschler et al argue that the polarization dependence is crucial to identifying
 gap structures on different Fermi surface pockets. As the temperature is lowered below
$T_c$, one expects the gap in the electronic spectrum to begin to
influence the scattering, provided the polarization states are
sensitive to the states being gapped.  The $A_{1g}$ configuration
samples the entire Brillouin zone, but the contributions from the
electron pockets are strongly diminished by screening effects;
therefore this polarization is primarily sensitive to the hole
pockets.  The major change with temperature in this channel occurs
near 100 cm$^{-1}$, which the authors identify with twice the hole
pocket maximum gap\cite{raman_theory} $\Delta_{max}$. On the other
hand, the data do not appear to be good enough at low energies to
allow statements about subgap excitations to be made on these
pockets.

\begin{figure*}[t]
\includegraphics[width=\textwidth]{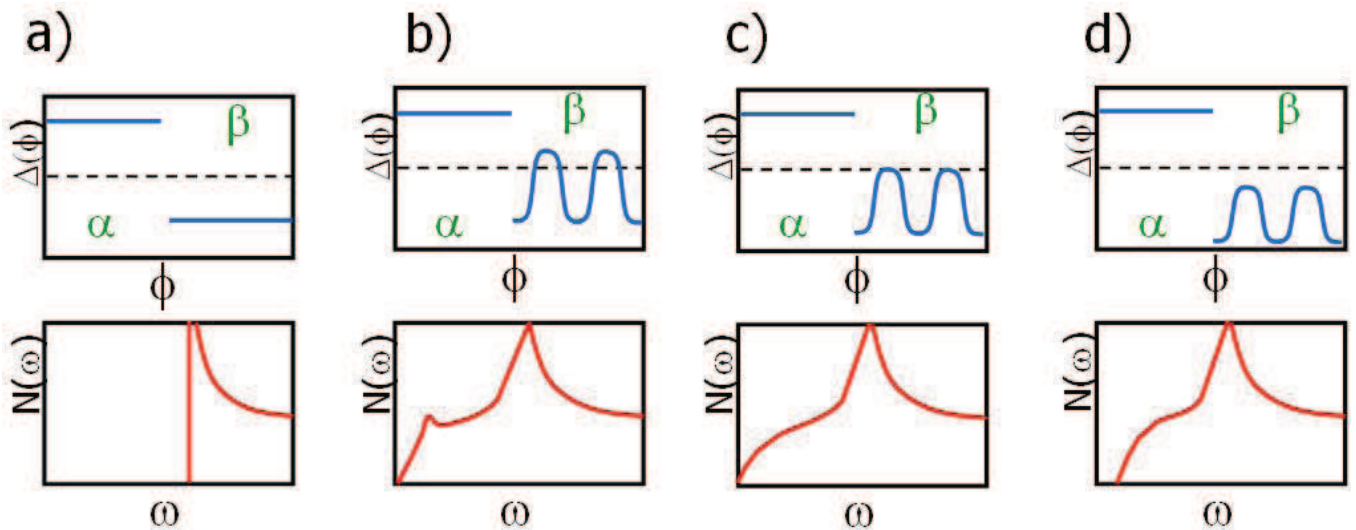}
\vspace{0cm} \caption{Schematic representations of possible
A$_{1g}$ type states discussed by Muschler et al.  Top panel:
order parameter plotted as function of local angle circling the
hole ($\alpha$) and electron ($\beta$) pockets.  Dashed line
represents $\Delta=0$. Bottom panel; corresponding density of
states for clean system. a) Isotropic $s_\pm$ state; b)
Anisotropic $s$-state with nodes on $\beta$ pockets; c) Same as b)
but for state with marginal (``kissing") nodes; d) Same as b) but
for deep gap minima. } \label{Fig1}
\end{figure*}

In the $B_{1g}$ polarization,  the predominant weight factors
occur in portions of the Brillouin zone near the X points where,
from the point of view of electronic structure calculations, it
seems unlikely that Fermi surface crossings exist for an electron
doped system.  Thus the lack of temperature dependence observed,
with the exception of a prominent phonon peak, is understandable.

By contrast  only the electron pockets
are sampled in the $B_{2g}$ polarization; here a much stronger temperature change is observed
as one enters the superconducting state, with a significant peak
near 70 cm$^{-1}$, implying a $\Delta_{max}$ of about 35 cm$^{-1}$
on these pockets.  Note  that these gaps are quite close to those
determined by ARPES measurements on a Ba-122 sample doped with 7\%
Co\cite{Terashima}. In addition, there is a clear low-$\omega$
power law in energy close to $\omega^{1/2}$ visible in the
low-temperature data for the optimal Co concentration sample. This
is the power law in the density of states one expects for an order
parameter on the electron pockets which barely touches the Fermi
surface (``kissing state", see Fig. 1c). It is not consistent with
a generic impurity band in an isotropic $s_\pm$ state.   While the
existence of nodes in the $s$-wave channel is ``accidental",
meaning it is determined by details of the pairing interaction
rather than by symmetry, theoretical calculations have indeed
found that the order parameter on the electron pockets comes quite
close to ``kissing" the Fermi surface\cite{sf_theory}, or slightly
overlapping (Fig. 1b-d), and thermal conductivity experiments on
the same material\cite{thermal_cond} have also been interpreted as
implying near-``kissing" states\cite{Mishra_kappa}. Addition of a
small amount of Co was found by Muschler et al. to lead to a small
range of energies where no excitations were visible, i.e. a small
gap of order 10 cm$^{-1}$. Were the effect of the additional Co
simply to add disorder to the system, this result would be
consistent with the suggestion by Mishra et al.\cite{nodelifting}
that intraband disorder scattering has a tendency to average an
extended $s$-wave gap with accidental nodes so as to eventually
``lift"  the nodes and create a full spectral gap. Some evidence
for the disorder interpretation is provided by the fact that the
$B_{2g}$ peak is considerably broader in the higher Co
concentration sample. However, more work needs to be done to rule
out a direct effect of the Co on the pair interaction itself via
doping or local structural modulations.


Why is the temperature dependence so much more significant on the
electron pockets?  At the transition, the hole pocket scattering
rate--given roughly by the position of the maximum in the Raman
intensity--appears to be many times the critical temperature for
the hole pockets. On the other hand, the analogous peak appears to
occur at an
energy  of order  about 20 cm$^{-1}$ on the electron pockets.  Why the
normal state lifetimes of electrons should be so much longer than
those of holes is not currently understood, but this result
appears to be consistent with transport\cite{MazinHu}
measurements. The ability to observe sharp transitions at low
temperature in the $B_{2g}$ channel implies further that the
electron pocket relaxation rate must be even smaller at low
temperatures; this is consistent with the collapse of the
relaxation rate as the gap opens, as observed in thermal
conductivity measurements\cite{Ong}.

The Muschler et al work provides internally consistent evidence
for order parameter nodes on the electron pockets, and for a
strong scattering rate anisotropy, largest on the hole pockets in
near-optimally Co doped Ba-122. This underlines the question of
why many other experiments appear to observe fully gapped states.
One possibility is that changes in electronic structure
 create more isotropic pairing states.  It appears
likely that the electronic structure of these materials can be
quite sensitive to materials parameters, and that relatively small
changes can rapidly tune the ``accidental" nodes away.  The exact
nature of this sensitivity will be interesting to try to sort out
in the future. Even if this view is correct, however, there is
mounting evidence that ARPES measures isotropic gaps even if the
bulk states probed by other experiments indicate nodes, implying a
possible strong dependence of this pairing state on surface
conditions as well. As in previous attempts to determine order
parameter symmetry in new superconducting materials, only by
comparing different experiments on high quality samples of various
materials may one expect a consensus to emerge.

\noindent
{}

\vspace{5mm}
\noindent

\noindent


\end{document}